\documentclass[aps,twocolumn]{revtex4}

\usepackage{graphicx}

\begin{document}

\title{Level density of $^{56}$Fe and low-energy enhancement of $\gamma$-strength function}
\author{A.V.~Voinov $^{1}$\footnote{Electronic address: voinov@ohio.edu},
S.M.~Grimes$^1$, U.~Agvaanluvsan$^2$, E.~Algin$^3$, T.~Belgya$^4$,
C.R.~Brune$^1$, M.~Guttormsen$^5$, M.J.~Hornish$^1$, T.~Massey$^1$,
G.E.~Mitchell$^{6,7}$, J.~Rekstad$^{5}$, A.~Schiller$^{8}$, S.~Siem$^{5}$}
\affiliation{$^1$ Department of Physics and Astronomy, Ohio University, Athens,
OH 45701, USA}
\affiliation{$^2$ Lawrence Livermore National Laboratory, L-414, 7000 East
Avenue, Livermore, CA 94551, USA}
\affiliation{$^3$ Department of Physics, Osmangazi University, Meselik,
Eskisehir, 26480 Turkey}
\affiliation{$^4$ Institute of Isotope and Surface Chemistry, Chemical Research
Centre HAS, P.O.Box 77, H-1525 Budapest, Hungary }
\affiliation{$^5$ Department of Physics, University of Oslo, N-0316 Oslo,
Norway}
\affiliation{$^6$ North Carolina State University, Raleigh, NC 27695, USA}
\affiliation{$^7$ Triangle Universities Nuclear Laboratory, Durham, NC 27708,
USA}
\affiliation{$^8$ NSCL, Michigan State University, East Lansing, MI 48824, USA}

\begin{abstract}
The $^{55}$Mn$(d,n)^{56}$Fe differential cross section is measured at
$E_d=7$~MeV\@. The $^{56}$Fe level density obtained from neutron evaporation
spectra is compared to the level density extracted from the
$^{57}$Fe$(^3$He,$\alpha\gamma)^{56}$Fe reaction by the Oslo-type technique.
Good agreement is found between the level densities determined by the two
methods. With the level density function obtained from the neutron evaporation
spectra, the $^{56}$Fe $\gamma$-strength function is also determined from the
first-generation $\gamma$ matrix of the Oslo experiment. The good agreement
between the past and present results for the $\gamma$-strength function
supports the validity of both methods and is consistent with the low-energy
enhancement of the $\gamma$ strength below $\sim 4$~MeV first discovered by the
Oslo method in iron and molybdenum isotopes.
\end{abstract}

\maketitle

\section{Introduction}

The unusual low-energy ($E_\gamma\lesssim 4$~MeV) enhancement of the
$\gamma$-strength function has been found recently -- first for $^{56}$Fe and
$^{57}$Fe nuclei \cite{fe,em} and then for the set of molybdenum isotopes
\cite{mo}\@. The $(^3$He,$\alpha\gamma)$ and $(^3$He,$^3$He$^\prime\gamma)$
reactions and sequential extraction procedure developed at the Oslo Cyclotron
Laboratory were used for this purpose. These results contradict the existing
understanding based on different extrapolations of the giant dipole resonance
(GDR) to the low-energy region. The strength of the observed enhancement may
indicate a different mechanism (in contrast to the GDR) in the low-energy
region of the $\gamma$-strength function. This requires additional theoretical
efforts to explain the observed enhancement.

The Oslo method allows one to extract both the nuclear level density (NLD) and
the $\gamma$-strength functions from the first-generation $\gamma$ matrix
$P(E_i,E_\gamma)$ obtained from particle-$\gamma$ coincidences in the
$(^3$He,$\alpha\gamma)$ reaction. Although it has been established that the
method works reasonably well in practice, the question of the applicability of
the Axel-Brink hypothesis remains open. This hypothesis assumes that the
$\gamma$-strength function depends only on the energy of the $\gamma$
transition and not on the excitation energies of the initial $E_i$ and final
$(E_i-E_\gamma)$ states \cite{axel,brink}. This assumption leads to the
factorization of the first-generation $\gamma$ matrix obtained from an Oslo
experiment as:
\begin{equation}
P(E_i,E_\gamma)\propto\rho(E_i-E_\gamma){\mathcal{T}}(E_\gamma),
\label{eq:axel}
\end{equation}
where $\rho$ is the NLD and ${\mathcal{T}}$ is the radiative transmission
coefficient, which is connected to the $\gamma$-strength function through the
relation:
$\mathcal{T}(E_\gamma)=2\pi\sum_{XL}f_{XL}(E_\gamma)E_\gamma^{(2L+1)}$\@. The
$\rho$ and ${\mathcal{T}}$ functions are determined by an iterative procedure
\cite{schi0} through the adjustment of these two functions at each data point
until a global $\chi^2$ minimum with the experimental $P(E_i,E_\gamma)$ matrix
is reached. Another assumption is that the $\gamma$ transitions originating
from some energy interval feed levels with the same decay properties as those
populated in the $(^3$He$,\alpha)$ reaction at the same excitation energy in
the residual nucleus. This assumption has been partially supported by
comparison of the results from two different reactions, namely
$(^3$He,$\alpha)$ and $(^3$He,$^3$He$^\prime$) \cite{Osloprove} populating the
same residual nucleus. Although the Oslo method has been thoroughly tested,
concern remains particularly about the validity of the Axel-Brink hypothesis.
For example, the theory developed in Ref.\ \cite{KMF} claims that the
$\gamma$ strength for spherical nuclei should depend on the temperature of the
final state, implying that the Axel-Brink hypothesis is not valid.

In order to address the above concerns, the NLD in Eq.\ (\ref{eq:axel}) should
be measured independently by a different kind of experiment. One of the most
reliable methods to extract the NLD below the particle separation threshold is
based on measurement of particle evaporation spectra from nuclear reactions.
Such spectra are described by a simple model of nuclear reactions based on
Hauser-Feshbach formalism; according to this formalism the shape of the
particle spectra depends only on the NLD of the final nuclei and the
transmission coefficients of outgoing particles. Because transmission
coefficients can be tested experimentally through the capture cross section of
an inverse reaction, the NLD can be deduced from spectra. The concern with this
method is with the possible contribution of pre-equilibrium and/or direct
reaction mechanisms which potentially can distort the shape of statistical
evaporation spectra and thus affect the NLD functions. The Oslo method uses
$\gamma$ transitions which have been proven to be statistical. Therefore the
comparison of NLDs obtained from these two experiments employing different
reactions and extraction procedures will not only enable the comparison of the
two methods, but also to estimate the possible distortion of particle
evaporation spectra caused by direct reaction contributions.

In this work the $^{55}$Mn$(d,n)^{56}$Fe reaction was investigated. The NLD
obtained from the neutron evaporation spectrum was analyzed and compared to the
NLD determined from the Oslo experiment, where the
$^{57}$Fe$(^3$He,$\alpha\gamma)^{56}$Fe reaction was used. Since the method and
results of the latter experiment have been thoroughly described in recent
publications \cite{fe,femo}, we concentrate on the description of the
$^{55}$Mn$(d,n)^{56}$Fe experiment.


\section{Experiment and method}

\begin{figure}[htb!]
\includegraphics[width=8.5cm]{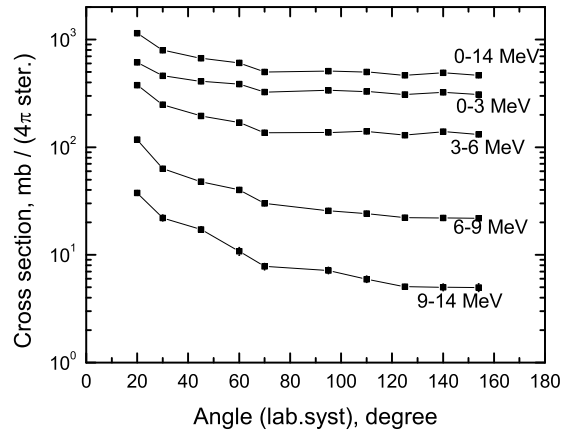}
\caption{The angular distribution for different energy groups of
outgoing neutrons from $^{55}$Mn$(d,n)$ reactions.}
\label{fig:fig1}
\end{figure}

\begin{figure}[htb!]
\includegraphics[width=8.5cm]{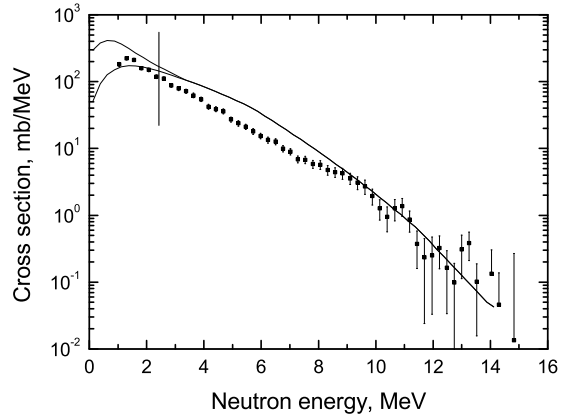}
\caption{The neutron evaporation spectrum at backward angles (points).
  The spectrum calculated with the Empire program taking into account the
  level density according to the microscopic model of Ref.~\cite{Gorely}(solid line).
   The vertical line shows the neutron energy limit beyond which the
first stage contribution to the total yield is greater than 90\%\@.}
\label{fig:fig2}
\end{figure}

The experiment was performed with a 7~MeV deuteron beam from the John Edwards
Accelerator Laboratory tandem at Ohio University. To measure the neutron
spectrum, the beam swinger facility \cite{americo} has been used. This allows
the measurement of angular distributions by rotating the incoming beam and the
target chamber with respect to the direction of outgoing neutrons. A self
supporting 0.74~mg manganese foil was used as a target. The energy of the
outgoing neutrons is determined by the time-of-flight method with a 7~m flight
path and NE213 neutron detectors. A 3~ns pulse width provided an energy
resolution of about 100~keV and 800~keV for 1 and 14~MeV neutrons,
respectively. The neutron detector efficiency was determined with the
calibrated neutron flux from the $^{27}$Al$(d,n)$ reaction on a thick Al target
at $E_d=7.44$~MeV \cite{Aleff}\@. This allowed the determination of the
detector efficiency from 0.2 to 14.5 MeV neutrons with an accuracy of
$\sim 6$\%\@. The neutron spectra were measured at nine different angles from
20 to 150 degrees to determine the angular distribution of outgoing neutrons.
Additional measurements with an empty target were performed at each angle in
order to determine the background contribution. The absolute cross section was
determined by taking into account the target thickness, the accumulated charge
of the incoming deuterons and the detector efficiency. The angular distribution
of outgoing neutrons is shown in Fig.\ \ref{fig:fig1}\@. The observed angular
anisotropy indicates the contribution from non-compound reactions at angles
less than $\sim 70^\circ$\@. The contribution of the non-isotropic part to the
total neutron cross section is estimated to be about 30\%\@. The deuteron
break-up mechanism is also responsible for the cross section anisotropy at
lower neutron energies. The angular dependence of the cross section at backward
angles is flat and assumed to be due to the compound nuclear mechanism.
Therefore the spectra averaged over backward angles have been used to extract
the NLD of the residual $^{56}$Fe nucleus (see Fig.\
\ref{fig:fig2})\@.

The procedure used to extract the NLD from the evaporation spectra was proposed
in  Ref.\ \cite{Vonach}\@. This extraction procedure is based on the
Hauser-Feshbach theory of compound nuclear reactions, according to which the
particle emission cross section is:
\begin{eqnarray}
\lefteqn{\frac{d\sigma}{d\varepsilon_b}(\varepsilon_a,\varepsilon_b)=}\\
\label{eq:HF}
&&\sum_{J\pi}\sigma^{\mathrm{CN}}(\varepsilon_a)\,\frac{\sum_{I\pi}
\Gamma_b(U,J,\pi,E,I,\pi)\rho_b(E,I,\pi)}{\Gamma(U,J,\pi)}\nonumber
\end{eqnarray}
with
\begin{eqnarray}
\lefteqn{\Gamma(U,J,\pi)=\sum_{b^\prime}\left(\sum_k
\Gamma_{b^\prime}(U,J,\pi,E_k,I_k,\pi_k)\,+\right.}\\
&&\left.\sum_{I^\prime\pi^\prime}\int_{E_c}^{U-B_{b^\prime}}dE^\prime\,
\Gamma_{b^\prime}(U,J,\pi,E^\prime,I^\prime,\pi^\prime)\,
\rho_{b^\prime}(E^\prime,I^\prime,\pi^\prime)\right).\nonumber
\end{eqnarray}
Here $\sigma^{CN}(\varepsilon_a)$ is the fusion cross section, $\varepsilon_a$
and $\varepsilon_b$ are energies of relative motion for incoming and outgoing
channels ($\varepsilon_b=U-E_k-B_b$, where $B_b$ is the separation energy of particle $b$
from the compound nucleus), the $\Gamma_{b}$
are the transmission coefficients of the outgoing particle, and the quantities
$(U,J,\pi)$ and $(E,I,\pi)$ are the energy, angular momentum, and parity of the
compound and residual nuclei, respectively, $E_c$ is the continuum edge. It follows from Eq.\ (\ref{eq:HF})
that the shape of the NLD is determined by both the transmission coefficients
of outgoing particles and the NLD of the residual nucleus $\rho_b(E,I,\pi)$\@.
Transmission coefficients can be calculated from optical model potentials
usually based on experimental data for elastic scattering and total cross
sections in the corresponding outgoing channel. This leaves the NLD as the only
unknown parameter, which can be extracted from this equation by using the
experimental differential cross section. Details and assumptions of this
procedure are described in Refs.\ \cite{Vonach,Vonach1}\@.

Neutron transmission coefficients are calculated from the optical model
potentials taken from the RIPL-2 data base \cite{RIPL}\@. Ten potentials have
been tested. These are potentials based on global systematics such as given by
Wilmore and Hodgson \cite{Wilmore} as well as those obtained for the local mass
range near $A=56$ nuclei. All of them have been found to give the same result
(the same shape of neutron evaporation spectra) within $\sim$10\% for
1--15~MeV\@. Finally, the potential of Wilmore and Hodgson has been adopted and
$10$\% errors are added to the neutron transmission coefficients. In order to
extract the NLD of the residual nucleus the following procedure was adopted:
({\it i}) The NLD model is chosen to calculate the differential cross section
of Eq.\ (\ref{eq:HF})\@. The parameters of the model were adjusted to reproduce
the experimental spectrum as closely as possible. ({\it ii}) The input NLD was
improved by binwise renormalization according to the expression:
\begin{equation}
\rho_b(E,I,\pi)=\rho_b(E,I,\pi)_{\mathrm{input}}
\frac{(d\sigma/d\varepsilon_b)_{\mathrm{meas}}}
{(d\sigma/d\varepsilon_b)_{\mathrm{calc}}}.
\label{eq:rho}
\end{equation}
The procedure described above is only correct when the main contribution to the
differential cross section comes from the first stage of the nuclear reaction
populating the residual nucleus of interest. In our case the second stage
contaminations open up above particle separation energies and come mainly from
$(d,\alpha n)$, $(d,pn)$ and $(d,nn)$ reactions. But as proposed in  Ref.\
\cite{chak}, as long as these contributions are less than the total error of
the extracted NLD, the energy interval chosen for the extraction of the NLD can
be extended beyond the particle separation threshold. Assuming 10\%
experimental errors (or more), the excitation energy interval where the second
stage contributions do not exceed 10\% is 0--11.5~MeV for the $^{55}$Mn$(d,xn)$
reaction (see Fig.\ \ref{fig:fig2})\@.


\section{Level density of $^{56}\textup\bf Fe$}

\begin{figure}[htb!]
\includegraphics[width=8.5cm]{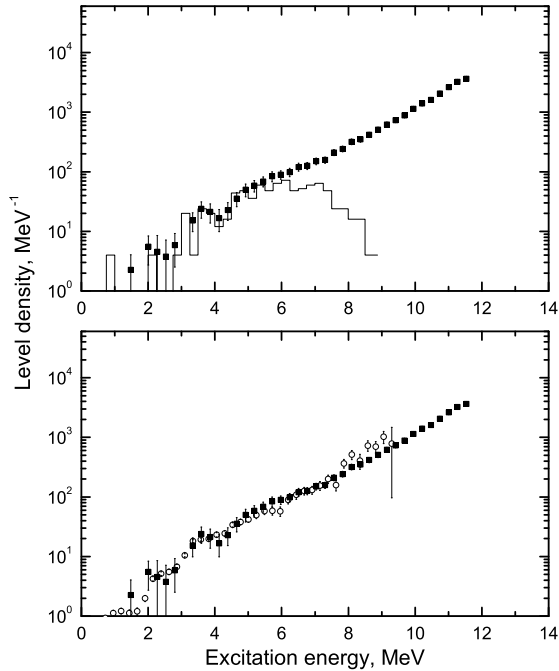}
\caption{The comparison of the NLD extracted from neutron evaporation spectra
(full circles) with discrete NLD (upper panel) and with NLD (open circles)
obtained from Oslo-type experiment (lower panel)\@.}
\label{fig:fig4}
\end{figure}

The extracted NLD for the $^{56}$Fe nucleus is shown in Fig.\ \ref{fig:fig4}
along with the density of discrete low-lying levels (upper panel) as well as
with the NLD obtained from the Oslo experiment (lower panel)\@. The good
counting statistics in the region corresponding to the location of known
discrete levels allowed an absolute normalization of the extracted NLD\@. This
is necessary because the scaling factor also depends on NLDs and transmission
coefficients of other outgoing channels (mainly proton and $\alpha$)\@.

Figure \ref{fig:fig4} demonstrates that up to about 6~MeV excitation energy the
extracted NLD almost perfectly follows the shape of the NLD function based on
discrete levels. This implies that the transmission coefficients used in
calculating the theoretical spectrum are correct. Above this point the discrete
levels are not complete and their density drops, while the NLD obtained from
our experiment continues to increase. It also agrees well with the NLD obtained
from the Oslo experiment \cite{femo}\@. One can see the good general agreement
up to $\sim 8$~MeV\@.

Both curves show the same step structures at around 4 and probably at 6~MeV
excitation energy. Similar steps have been interpreted in \cite{femo} as a
result of the breaking of Cooper pairs. Above about $\sim 7.8$~MeV these curves
start to diverge, ending with differences of about 50\% at an excitation energy
of $\sim 8.8$~MeV\@. This deviation may stem from imperfections of the two
methods, perhaps connected with violations of the underlying assumptions. Some
local deviations of the neutron transmission coefficients are also possible.
However, the almost perfect agreement below $\sim8$~MeV gives confidence in
both methods.

Generally, one can conclude that in spite of the fact that these two methods
use different underlying assumptions, different nuclear reactions, and
different mathematical techniques to extract NLD, a fairly consistent result
has been obtained. This implies that the statistical mechanism dominates in
both reactions. The observed difference between NLDs at higher excitation
energy requires further investigation. The observed small bump in the
differential cross section at $\sim 9$~MeV of the neutron energy transforms to
the corresponding bump in the extracted NLD at about 5.5~MeV of excitation
energy. Previously, similar structures were observed in Oslo type experiments
for a variety of nuclei and have been explained as a result of pairing
correlations. The NLD of $^{56}$Fe from the Oslo experiment also exhibits
structure at $\sim 5.5$~MeV, although the shape is slightly different from that
observed in the particle evaporation spectrum in Fig.\ \ref{fig:fig4}\@. This
is probably due to systematic uncertainties still inherent in  both methods.

\begin{figure}[htb!]
\includegraphics[width=7.3cm]{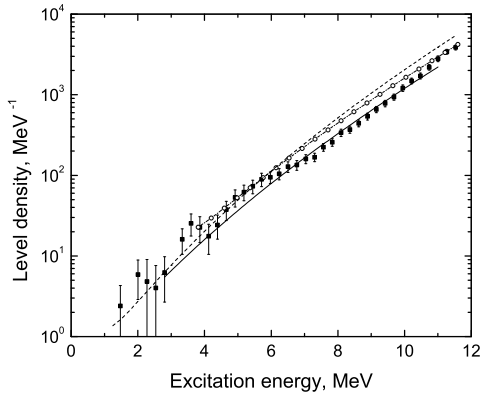}
\caption{The NLD extracted from neutron evaporation spectra (full circles)
compared to Fermi-gas (dashed) and microscopical model (open circles)
calculations. The full line shows the fit to the data.}
\label{fig:fig5}
\end{figure}

The presence of the step structure in the NLD is an important feature from a
practical point of view, because this might introduce corrections to available
systematics of NLD parameters widely used for calculation of reaction rates.
Almost all systematics are primarily based on NLDs obtained from neutron
resonance spacings at the neutron separation energy $B_n$ and the NLD of
low-lying discrete levels. To calculate the total NLD at $B_n$ the spin cutoff
parameter is used; this parameter is not known to high accuracy in this energy
region. The NLD in the intermediate region is often obtained by a simple
interpolation between these two anchor points, assuming some NLD model. The
Fermi-gas model is most often used. For example the NLD according to the
Fermi-gas model with parameters from the latest systematics of Ref.\
\cite{Egidy} is shown in Fig.\ \ref{fig:fig5}\@. This curve fits the discrete
NLD well, but overestimates the experimental points by a factor of $\sim$~1.7
at higher excitation energies. A slightly better result is given by the
microscopic model \cite{Gorely} recommended in the RIPL-2 data base
\cite{RIPL}\@. However, because this NLD is also renormalized to the neutron
resonance spacing and density of low-lying levels, the discrepancy is still
present. These or other similar models may cause a sizable overestimation of
calculated reaction cross sections and incorrect ratios of cross sections for
different channels.

To fit the experimental NLD the Fermi-gas formula has been adopted:
\begin{equation}
\rho(E)=\frac{\exp(2\sqrt{a(E-\delta)})}
{12\sqrt{2}a^{1/4}(E-\delta)^{5/4}\sigma},
\end{equation}
where the spin cutoff parameter $\sigma$ based on a rigid body moment of
inertia is expressed as:
\begin{equation}
\sigma^2=0.0145A^{5/3}\sqrt{(E-\delta)/a}.
\end{equation}
The Fermi-gas parameters which give the best fit to experimental points between
2 and 11 MeV are $a=5.65(10)$~MeV$^{-1}$ and $\delta=0.65(20)$~MeV\@. The spin
cutoff formula based on half of the moment of inertia does not cause large
changes. These parameters are slightly smaller than those obtained from the
systematics of Ref.\ \cite{Egidy} ($a=6.16$~MeV$^{-1}$, $\delta=0.8$~MeV)\@.
However, this difference is enough to cause $\sim 70\%$ discrepancy in
corresponding NLDs at 6--10~MeV excitation energies (see Fig.\
\ref{fig:fig5})\@.

\begin{table}[htb!]
\caption{NLD parameters of $^{56}$Fe from different systematics.}
\begin{tabular}{l|cc}\hline\hline
Systematics     &$a$ &$\delta$\\\hline
T.~von~Egidy    &6.16&0.8     \\
G.~Rohr         &5.61&2.81    \\
S.I.~Al-Quraishi&5.98&1.68    \\
Present         &5.65(10)&0.65(20)    \\\hline\hline
\end{tabular}
\end{table}

The NLD parameters for $^{56}$Fe have also been obtained in Ref.\ \cite{Lu} and
\cite{sprinzak} from $(p,p^\prime)$, $(p,\alpha)$ and $(\alpha,\alpha^\prime)$
reactions. Despite the fact that the same reactions and the same techniques
have been used in these experiments, different results were obtained. The first
measurement reported $a=5.7$~MeV$^{-1}$ and $\delta=0.7$~MeV (in good agreement
with our results), while the second measurement gives $a=6.5$~MeV$^{-1}$ from
$(p,p^\prime)$ and $a=7.0$~MeV$^{-1}$ from $(p,\alpha)$ reactions. Part of the
problem might be the strong correlation of the $a$ and $\delta$ parameters when
they are extracted from particle evaporation spectra. In the case when the
extracted NLD is not normalized, one needs to assume some value of $\delta$ in
order to obtain the value of $a$\@. As has been shown in \cite{Lu}, a 1~MeV
increase of $\delta$ causes approximately 1~MeV$^{-1}$ decrease of parameter
$a$ (the correlation is negative)\@.

More recently V.~Mishra {\sl et al}.\ \cite{Mishra} reexamined this problem.
Their result confirmed the conclusion of \cite{Lu} if it is assumed that the
slope of the NLD is matched. On the other hand, if the parameters are extracted
from absolute value of NLD at a particular point (as in neutron resonance
analysis), the correlation between $a$ and $\delta$ has positive sign.

In addition to the von Egidy compilation, other compilations have been prepared
by Rohr \cite{Rohr} and Al-Quraishi {\sl et al}.\ \cite{Quraishi}\@. These
parameters are shown in Table I\@. It is interesting to note that the three
compilations have predicted slopes which are within 10\% of one another (same
nuclear temperature)\@. The magnitude predicted for the NLD between 6 and 10
MeV differs by more than an order of magnitude. Rohr predicts the smallest NLD,
Al-Quraishi next highest, the present data are second largest and von Egidy
{\sl et al}.\ predict the highest NLD\@. These differences in magnitude point
out the importance of performing absolute magnitude normalization in inferring
NLD\@. In this work, we were able to obtain the absolute NLD function by
normalizing it to the NLD of known low-lying levels. This gives us the
possibility of obtaining both the $a$ and $\delta$ values simultaneously.

Thus, in order to establish the systematics of NLD parameters, it is necessary
to take into account the absolute NLD for the whole excitation energy interval.
Such information can be obtained either from particle evaporation spectra by
using Eqs.\ (\ref{eq:HF}) and (\ref{eq:rho}) or from the Oslo type experiments
by using the sequential extraction method applied to the particle-$\gamma$
coincidence matrix.

\section{$\gamma$-strength function in $^{56}\bf\textup Fe$}

The $\gamma$-strength functions for $^{56}$Fe and $^{57}$Fe have been obtained
from an Oslo type experiment in Ref.\ \cite{fe}\@. The striking feature of
these functions is the increase of $\gamma$ strength in the region below 4~MeV;
this cannot be understood within existing models. The main concern about the
Oslo method is that possible violation of the Axel-Brink hypothesis might
result in some systematic deviation of the obtained $\gamma$ strength -- that
is, if the $\gamma$-strength function depends not only on the energy of the
$\gamma$ transition, but also on the temperature of the final state. Such
temperature dependence may stem from the temperature dependence of the GDR
width caused by different damping mechanisms debated in the literature
\cite{damping}\@. According to the Fermi-liquid model \cite{FL} the width is
determined by the collision of quasi particles in the nuclear volume that
results in the following temperature dependence:
\begin{equation}
\Gamma(E_\gamma,T) \propto (E_\gamma^2+4\pi^2T^2),
\label{eq:temp}
\end{equation}
where $T=\sqrt{U/a)}$\@. But as has been shown in Ref.\ \cite{Dyaper}, the
systematic deviation of the $\gamma$-strength function due to such temperature
dependence is about only 15\% in the region below 2~MeV\@. The other mechanism
is connected to nuclear shape fluctuations, leading to a square root
temperature dependence of the GDR width. The experiments on Sn and Pb
\cite{Sn_temp} support the temperature dependence of the GDR, but the mechanism
responsible for such effects is still under debate. A good fit to these
experimental data is obtained with the Fermi-liquid model accounting for
damping of the GDR according to Eq.\ (\ref{eq:temp}), while taking into account
the dipole-quadrupole interaction term arising from the nuclear deformation
\cite{Mugh}\@.

\begin{figure}[t!]
\includegraphics[totalheight=7.0cm]{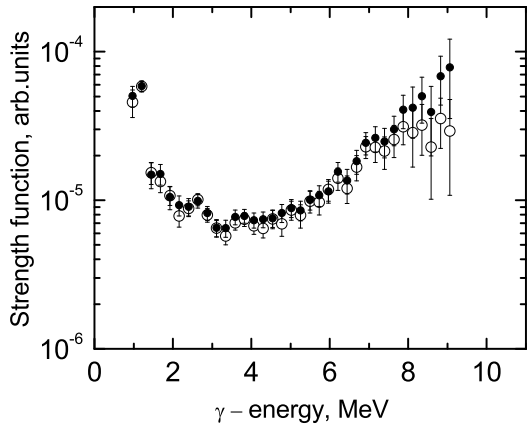}
\caption{The $\gamma$-strength function obtained from Oslo first-generation
matrix $P(E_\gamma,E_x)$ with NLD from $(d,n)$ reaction (filled circles)\@. The
$\gamma$-strength function obtained solely from $P(E_\gamma,E_x)$ (open
circles)\@.}
\label{fig:fig6}
\end{figure}

The low-energy enhancement of the $\gamma$-strength function observed in the
Oslo experiment for iron isotopes might stem from different (i.e.\ not GDR)
modes of nuclear excitation and therefore might exhibit a different temperature
dependence. Therefore it is important to investigate the validity of the
Axel-Brink hypothesis in the Oslo method for the iron isotopes. For this
purpose we can use the first generation matrix $P(E_i,E_\gamma)$ obtained from
the Oslo $^{57}$Fe$(^3$He,$\alpha\gamma)^{56}$Fe experiment and the NLD from
$^{55}$Mn$(d,n)^{56}$Fe reaction to obtain the $\gamma$-strength function of
the $^{56}$Fe, giving
\begin{equation}
f(E_\gamma,E_i)=\frac{1}{2\pi}\frac{N(E_i)P(E_i,E_\gamma)}
{\rho(E_i-E_\gamma)E_\gamma^3},
\label{eq:rsf}
\end{equation}
where $E_i=E_\gamma+E_f$\@. It is clear that the obtained $\gamma$-strength
function should not deviate considerably from that extracted solely from the
$P(E_\gamma,E_x)$ obtained by the sequential iterative procedure because the
corresponding NLD functions agree well (see Fig.\ \ref{fig:fig4})\@. However,
due to slightly different slopes of these functions in the region of 4--7~MeV
and $\sim 1.7$ times disagreement above this region (which in principle can be
caused by temperature effects), it is interesting to investigate the magnitude
of corresponding local changes in the $\gamma$-strength function. The
normalization constant $N(E_i)$ in Eq.\ (\ref{eq:rsf}) has been determined in
Ref.\ \cite{mo} at $E_i=B_n$ by using supplementary experimental information
from systematics of total $\gamma$ widths of initial states (neutron
resonances)\@. The comparison of $\gamma$-strength functions obtained from both
the sequential extraction Oslo method and Eq.\ (\ref{eq:rsf}) using the NLD
from the evaporation spectrum is shown in Fig.\ \ref{fig:fig6}\@. There is no
significant disagreement. It can be concluded that possible temperature effects
on the extracted $\gamma$-strength function are rather small compared to total
uncertainties in the experimental data. Thus the applicability of the
Axel-Brink hypothesis in the Oslo method is justified within the accuracy of
the experimental data. However, one should keep in mind that in both cases
investigated here, one extracts an averaged $\gamma$-strength function for a
wide region (several MeV) of final energies. In order to detect a possible
temperature dependence of the $\gamma$-strength function, one could extract the
$\gamma$-strength function from several limited regions of final energy,
however, for this, one needs to increase the statistics of the experiment to be
able to reduce statistical errors.

\section{Summary and conclusions}

The $^{56}$Fe NLD has been extracted from the neutron evaporation spectrum of
the $^{55}$Mn$(d,n)^{56}$Fe reaction. This NLD has been compared to that
obtained from the $^{57}$Fe$(^3$He,$\alpha\gamma)^{56}$Fe reaction by using the
sequential extraction technique developed at the Oslo Cyclotron Laboratory. The
NLDs obtained from these two different types of experiments are in good
agreement with each other. This indicates the consistency of these two methods
and possibility of applying such methods to investigation of a broader range of
nuclei. This agreement helps to eliminate most of the potential systematic
errors inherent in these methods, including such important problems as the
unknown contribution of direct processes in particle evaporation spectra. The
neutron evaporation spectrum from the $(d,n)$ reaction measured at backward
angles does not contain such a contribution.

The NLD function of $^{56}$Fe can be fit by the conventional Fermi-gas model in
the region of 2--11~MeV excitation energy. However, a local deviation of the
order of $\sim 40$\% has been observed at 5~MeV\@. The presence of this
structure leads to disagreement in the Fermi-gas parameters obtained from our
current data and from available systematics. In order to verify the character
of these structures, more experimental data are needed for neighboring nuclei.

The $\gamma$-strength function for the $^{56}$Fe isotope obtained in Ref.\
\cite{fe} has been extracted by using the NLD from the neutron evaporation
spectrum. The new $\gamma$-strength function agrees well with the previous one
within experimental errors. This indicates small temperature effects on the
extracted $\gamma$-strength function. Thus the Axel-Brink hypothesis used in
the Oslo method appears to be justified within the present experimental
uncertainties.

\acknowledgments

The authors acknowledge support from National Nuclear Security Administration
under the Stewardship Science Academic Alliances program through DOE Research
Grant No.\ DE-FG03-03-NA0074 and DE-FG03-03-NA0076\@. Part of this work was
performed under the auspices of the U.S. Department of Energy by the University
of California, Lawrence Livermore National Laboratory under Contract
W-7405-ENG-48\@. Financial support from U.S. Department of Energy Grant No.\
DE-FG02-97-ER41042 and the Norwegian Research Council (NFR) is gratefully
acknowledged.

\end{document}